\documentclass[10pt, letterpaper]{emulateapj}

\begin{document}
 
\title{The orbits of the quadruple star system 88 Tau A from PHASES differential astrometry and radial velocity}

\author{Benjamin F.~Lane\altaffilmark{1}, Matthew W.~Muterspaugh\altaffilmark{2,3}, 
Francis C. Fekel\altaffilmark{4}, Michael Williamson\altaffilmark{4}, Stanley Browne\altaffilmark{2}, Maciej Konacki\altaffilmark{5}, 
Bernard F.~Burke\altaffilmark{1}, M.~M.~Colavita\altaffilmark{6}, S.~R.~Kulkarni\altaffilmark{7}, M.~Shao\altaffilmark{6}
}
\altaffiltext{1}{MIT Kavli Institute for Astrophysics and Space Research, MIT Department of Physics, 70 Vassar Street, Cambridge, MA 02139}
\altaffiltext{2}{University of California, Space Sciences Laboratory, 7 Gauss Way, Berkeley, CA 94720-7450}
\altaffiltext{3}{Townes Fellow}
\altaffiltext{4}{Center of Excellence in Information Systems, Tennessee State University, 3500 John A. Merritt Blvd, Box 9501, Nashville, TN 37209}
\altaffiltext{5}{Nicolaus Copernicus Astronomical Center, Polish Academy of Sciences, Rabianska 8, 87-100 Torun, Poland}
\altaffiltext{6}{Jet Propulsion Laboratory, California Institute of Technology, 4800 Oak Grove Dr., Pasadena, CA 91109}
\altaffiltext{7}{Division of Physics, Mathematics and Astronomy, 105-24, California Institute of Technology, Pasadena, CA 91125}

\begin{abstract} 
We have used high precision differential astrometry from the Palomar High-precision Astrometric Search for Exoplanet Systems (PHASES)
project and radial velocity measurements covering a time-span of 20 years to determine the 
orbital parameters of the 88 Tau A system. 88 Tau is a complex hierarchical multiple system comprising a total of six stars; we have studied
the brightest 4, consisting of two short-period pairs orbiting each other with an $\sim$18-year period.  We present 
the first orbital solution for one of the short-period pairs, and determine the masses of the components and 
distance to the system to the level of a few percent. In addition, our astrometric measurements allow us to make the 
first determination of  the mutual inclinations of the orbits. We find that the sub-systems are not coplanar.
\end{abstract}

\keywords{techniques:interferometric--star:88 Tau}
 
\section{Introduction}

\begin{figure}
\figurenum{1}
\plotone{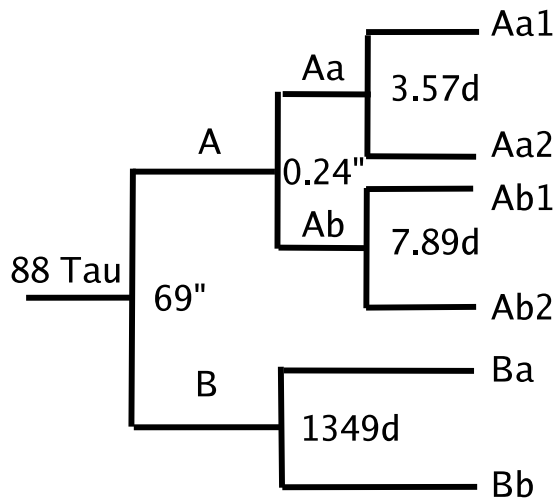}
\figcaption{A schematic diagram of the 88 Tau system. 
\label{fig1}}

\end{figure}

88 Tau (HD 29140, HR 1458, HIP 21402) is a bright 
($m_V=4.25, m_K=3.69\pm 0.25$; Skrutskie et al. 2006\nocite{2mass}), nearby
($\sim 50$pc) hierarchical sextuple stellar system \citep{msc}. The A
component contains a pair of systems (designated Aa
and Ab) in an $\sim18$-year \citep{balega99}
 orbit that has been resolved by speckle interferometry \citep{mac87}.
 The Aa component is a known spectroscopic binary system 
 ($P\sim 3.57$-day), with a composite
spectral type of A5m \citep{c69}. In previous work it had been noted \citep{bc88} 
that the A system is likely complex, with possibly as many as 5 components. 
\citet{balega99} noted a discrepancy between the total estimated 
mass of this system based on photometry and spectral types, and 
the total mass derived from the visual orbit and {\em Hipparcos}
parallax.  In this work we have determined that, like the Aa component, the Ab 
component is a double-lined binary; this newly-resolved binary has a period of 7.89 
days. Finally, there is a common-proper-motion companion, labeled B, located 
$\sim$69 arcseconds away from the A system;  it, too, is
known to be a binary \citep{tg01}.  For clarity we provide a 
schematic of this complex system in Figure \ref{fig1}.

There are several reasons why multiple stellar systems such as 88 Tau
merit attention: first, binary orbits make it possible to measure accurate
stellar masses and distances, while the larger number of
presumably co-eval stars allows one to impose the additional
constraint that any given model must accurately match all of the
stars. This approach has proven particularly fruitful when applied 
to another famous hierarchical sextuple system: Castor\citep{tr02}.
Second, as outlined in \citet{st02}, the relative orientations
of the orbital angular momenta allow one to constrain the properties
of the cloud from which the stars are thought to have formed, as well as the
subsequent dynamical decay process. Despite their value, observational
problems have limited the number of triple or higher-order systems
with accurately measured orbits to fewer than 10. Given their hierarchical nature it is
often the case that either the close system is unresolvable or the
outer system has an impractically long orbital period. 

With the advent of long-baseline stellar interferometry, and more recently 
phase-referenced long-baseline interferometric astrometry \citep{lm04}
capable of 10--20 $\mu$-arcsecond astrometric precision between 
pairs of stars with separations in the range 0.05--1 arcsecond, it
has become possible to resolve the orbital motion of several 
interesting multiple systems \citep{kapPeg,v819Her}. Here we 
report on astrometric and radial velocity measurements of the 88 Tau A
system, which allow us to constrain the orbits of the 3.57-day, 7.89-day and 
18-year components with improved precision, and for the first time 
provide a relative orientation of the orbits as well as component masses. 

Astrometric measurements were made with the Palomar Testbed
Interferometer \citep{colavita99} as part of the Palomar
High-precision Astrometric Search for Exoplanet Systems (PHASES)
program \citep{limits}.  The Palomar Testbed Interferometer is located
on Palomar Mountain near San Diego, CA. It was
developed by the Jet Propulsion Laboratory, California Institute of
Technology for NASA as a testbed for interferometric techniques
applicable to the Keck Interferometer and the
Space Interferometry Mission (SIM). It operates in the J (1.2 $\mu$m), H
(1.6 $\mu$m), and K (2.2 $\mu$m) bands and combines starlight from two out of
three available 40 cm apertures. The apertures form a triangle with 86
and 110 m baselines.

\section{Observations \& Models}

\subsection{PHASES Astrometry}
88 Tau A was successfuly observed with PTI on 29 nights in 2004--2007
with the use of the phase-referenced fringe-scanning mode \citep{lm04} developed
for high-precision astrometry; the data were reduced with the
algorithms described therein, as well as with the modifications
described in \citet{delEqu}. 

The obtained differential astrometry is listed in Table
\ref{tab:data}.  Note that the astrometry on any single night is
essentially that of a single-baseline interferometer, yielding a very
small error in the direction aligned with the baseline, but limited to
the effect of Earth-rotation synthesis in the perpendicular direction.
The median minor-axis formal uncertainty is 10 $\mu$arcseconds, while
the median major-axis uncertainty is 312 $\mu$arcseconds.  To
properly weight the data set when doing a combined fit with previous
astrometry and radial velocity data, we fit an orbital model to the
PHASES astrometry by itself, and rescaled the formal uncertainties so
as to yield a reduced $\chi^2$ of unity; the resulting scale factor
was 2.5, indicating a substantial amount of excess scatter beyond the
internal error estimates. We do not believe this scatter to be due to
the effect of starspots, given that the {\it Hipparcos} photometry
of this system indicates a scatter of no more than 5 mmag; the resulting maximum
starspot-induced astrometric noise would be $\sim 4 \mu$arcseconds
\citep{v819Her}.  We have however identified possible instrumental
sources of this systematic error and developed methods for reducing
it, see \cite{muOri}. Nevertheless, the existing astrometry is
sufficient to detect astrometric motion induced by the short-period subsystems. 

\begin{deluxetable*}{ccccccccccccl}
\tabletypesize{\footnotesize}
\tablecaption{PHASES Astrometric data for 88 Tau A\label{tab:data}}
\tablewidth{0pt}
\tablehead{
HJD-2400000.5 &  $\Delta$RA$\cos(\delta)$ & $\Delta$Dec & $\sigma_{min}$ & $\sigma_{maj}$ & $\phi_e$ & $\sigma_{RA}$ & $\sigma_{Dec}$ & $\frac{\sigma^2_{RA,Dec}}{\sigma_{RA}\sigma_{Dec}} $ & N & ADC & Align & Rate \\
 & (mas) & (mas) & ($\mu$as) & ($\mu$as) & (deg) & ($\mu$as) & ($\mu$as) & & & & &(Hz)
}
\startdata 
52979.34108 & -32.1751 & -101.5913 & 20.6 & 477.9 & 163.71 & 458.8 & 135.5 & -0.98733 & 2353  & 0 & 0 &100  \\
53034.13365 & -40.3179 & -91.5425 & 14.0 & 311.9 & 152.78 & 277.4 & 143.2 & -0.99394 & 2951  & 0 & 0 &100 \\
53250.50169 & -65.6825 & -53.5438 & 40.2 & 1775.4 & 147.24 & 1493.2 & 961.3 & -0.99876 & 1027 & 0 & 0 &100  \\
53271.46723 & -70.0370 & -48.2688 & 16.6 & 669.7 & 150.15 & 580.9 & 333.7 & -0.99836 & 4005 & 0 & 0 &100\\
53291.40191 & -71.5816 & -44.8765 & 88.3 & 1759.4 & 148.37 & 1498.7 & 925.8 & -0.99371 & 408  & 0 & 0 &100\\
53294.47661 & -71.0249 & -44.6002 & 32.3 & 1700.7 & 164.09 & 1635.6 & 467.2 & -0.99742 & 1355   & 0 & 0 &100\\
53312.38113 & -74.1285 & -40.9475 & 7.8 & 68.0 & 154.15 & 61.3 & 30.5 & -0.95879 & 8150 & 0 & 0 &100\\
53320.33408 & -74.5082 & -39.6052 & 40.3 & 2076.1 & 150.23 & 1802.1 & 1031.5 & -0.99899 & 1838  & 0 & 0 &100 \\
53340.29304 & -77.4848 & -35.5088 & 18.2 & 215.1 & 152.98 & 191.8 & 99.1 & -0.97862 & 3602   & 0 & 0 &100\\
53341.28228 & -77.5104 & -35.2746 & 17.2 & 551.5 & 150.70 & 481.0 & 270.3 & -0.99734 & 3630   & 0 & 0 &100\\
53605.52764 & -104.7426 & 15.6166 & 18.3 & 956.5 & 147.64 & 808.1 & 512.1 & -0.99911 & 2855   & 0 & 0 &100\\
53606.51559 & -106.5080 & 16.8975 & 15.7 & 972.4 & 146.14 & 807.5 & 542.0 & -0.99939 & 3205   & 0 & 0 &100\\
53614.50698 & -105.6799 & 17.1892 & 25.4 & 773.7 & 147.94 & 655.8 & 411.2 & -0.99735 & 1716   & 0 & 0 &100\\
53687.37864 & -112.2999 & 31.2082 & 24.7 & 227.4 & 31.80 & 193.7 & 121.7 & 0.97114 & 3260  & 0 & 0 &100 \\
53711.29715 & -114.6779 & 35.8970 & 80.7 & 880.5 & 29.41 & 768.1 & 438.0 & 0.97743 & 833   & 1 & 0 &100\\
53712.28437 & -114.0278 & 36.0764 & 41.2 & 326.1 & 27.87 & 288.9 & 156.8 & 0.95491 & 2889   & 1 & 0 &100\\
53789.14419 & -120.0382 & 51.3801 & 59.4 & 3107.9 & 40.17 & 2375.3 & 2005.2 & 0.99925 & 761  & 1 & 0 &100\\
53790.13775 & -119.0373 & 52.0709 & 77.5 & 2663.3 & 39.39 & 2058.8 & 1691.3 & 0.99824 & 624  & 1 & 0 & 100 \\
54030.46119 & -135.6408 & 94.2582 & 50.4 & 1514.6 & 163.75 & 1454.1 & 426.6 & -0.99240 & 1268  & 1 & 1 & 50 \\
54055.38631 & -136.2430 & 98.5452 & 48.0 & 870.1 & 161.74 & 826.4 & 276.5 & -0.98318 & 1679   & 1 & 1 & 50\\
54061.37247 & -136.6968 & 99.4972 & 16.3 & 302.1 & 163.08 & 289.0 & 89.3 & -0.98165 & 4216& 1 & 1 &50\\
54075.33188 & -136.8574 & 101.6204 & 13.6 & 250.1 & 162.18 & 238.2 & 77.7 & -0.98300 & 5902 & 1 & 1 &50\\
54083.30564 & -137.8524 & 103.1506 & 31.6 & 688.5 & 161.39 & 652.6 & 221.8 & -0.98863 & 3845  & 1 & 1 &50\\
54084.31595 & -136.7985 & 103.1084 & 63.9 & 1741.5 & 163.11 & 1666.5 & 509.6 & -0.99139 & 1182  & 1 & 1 &50 \\
54103.25456 & -137.6742 & 106.3338 & 42.7 & 1011.9 & 162.34 & 964.3 & 309.6 & -0.98949 & 743   & 1 & 1 &50\\
54138.15348 & -139.3256 & 112.2494 & 28.6 & 457.9 & 161.38 & 434.1 & 148.7 & -0.97916 & 3142  & 1 & 1 &50 \\
\enddata
\tablecomments{All quantities are in the ICRS 2000.0 reference frame. The uncertainty values 
presented in these data have been scaled by a factor of 2.5 over the formal (internal) uncertainties 
for each night.  Column 6 ($\phi_e$) is the angle between the major axis of the uncertainty 
ellipse and the right ascension axis, measured from increasing differential right ascension 
through increasing differential declination. N is the number of scans obtained in a night; each scan 
typically represents 0.5--1 second of integration. ADC indicates that the observations made use of the 
automatic dispersion compensator. Rate indicates the tracking rate of the fringe tracker used to stabilize the
fringe phase during measurement. Align indicates whether or not the automatic alignment 
system was used to stabilize the system pupil.  }
\end{deluxetable*} 

\subsection{Previous Astrometry}

In addition to our astrometry, 88 Tau A has been followed by a number of
observers with speckle-interferometric techniques. We use 20
observations tabulated in the {\em 4th Catalog of 
Interferometric Measurements of Binary Stars\footnote{\tt http://ad.usno.navy.mil/wds/int4.html} } \citep{int4}
to further constrain our fit. Although of somewhat lower precision, the considerable
time-baseline (including observations dating from 1985) help constrain
the parameters of the wide orbit. In many cases the published
astrometry lacks uncertainties, and we therefore assigned a 
plausible initial uncertainty of 3 mas in separation and 2 degrees in 
position angle to these points (we used published uncertainties for the 
points where such were available). We then performed a least-squares fit of a single Keplerian
orbital model (corresponding to the Aa-Ab orbit - the subsystems are far too small
to be detected by these data), and scaled all of the uncertainties
so as to yield a reduced $\chi^2$ of unity.  We find the average
uncertainty in separation to be 5 milli-arcseconds, and the average
position-angle uncertainty to be 3 degrees.

\subsection{Spectroscopic observations and reductions}

\begin{deluxetable*}{lcccccccc}
\tabletypesize{\scriptsize}
\tablecaption{KPNO Radial Velocity data for 88 Tau A\label{tab:dataknpo}}
\tablewidth{0pt}
\tablehead{
HJD-2400000.5 &  $V_{Aa1}$& Weight\tablenotemark{a} & $V_{Aa2}$ & Weight\tablenotemark{b} & $V_{Ab1}$ & Weight\tablenotemark{c} & $V_{Ab2}$ & Weight\tablenotemark{d} \\
                              &   (${\rm km\,s^{-1}}$) & & (${\rm km\,s^{-1}}$) & & (${\rm km\,s^{-1}}$) & & (${\rm km \,s^{-1}}$) & 
}
\startdata 
 45718.327 & 91.4 & 1.0  & -66.5 & 1.0 & - & - & - & - \\
 46388.389 & -48.7 & 1.0  & 141.6 & 1.0 & 40.1 & 1.0 & 1.1 & 1.0 \\
 46390.212 & 99.8 & 1.0  & -85.2 & 1.0 & -5.8 & 1.0 & - & - \\
 46718.473 & 96.5 & 1.0  & -93.7 & 1.0 & 59.8 & 0.4 & -4.5 & 1.0 \\
 46720.487 & -51.6 & 1.0  & 139.5 & 1.0 & - & - & - & - \\
 47152.190 & -45.3 & 1.0  & 123.6 & 1.0 & 57.9 & 1.0 & - & - \\
 47245.112 & -50.2 & 1.0  & 131.5 & 1.0 & - & - & - & - \\
 47456.264 & -52.2 & 1.0  & 129.7 & 1.0 & 9.3 & 1.0 & - & - \\
 47556.098 & -57.7 & 1.0  & 135.8 & 1.0 & - & - & - & - \\
 47624.100 & -53.5 & 1.0  & 125.7 & 1.0 & - & - & - & - \\
 47626.145 & 66.5 & 1.0  & -57.7 & 1.0 & - & - & 4.3 & 1.0 \\
 47627.123 & -49.3 & 1.0  & 118.9 & 1.0 & - & - & - & - \\
 48345.123 & -60.5 & 1.0  & 133.2 & 1.0 & - & - & - & - \\
 48347.103 & 96.5 & 1.0  & -102.4 & 1.0 & 9.4 & 1.0 & 69.9 & 0.4 \\
 48356.100 & -58.6 & 1.0  & 128.6 & 1.0 & 10.0 & 1.0 & 59.1 & 1.0 \\
 48505.518 & -46.5 & 1.0  & 106.1 & 1.0 & 14.0 & 1.0 & 53.0 & 1.0 \\
 48573.407 & -49.1 & 1.0  & 112.7 & 1.0 & 42.5 & 0.4 & 26.2 & 0.4 \\
 48604.284 & 90.2 & 1.0  & -98.1 & 1.0 & 57.7 & 0.4 & 15.4 & 1.0 \\
 48607.306 & 75.8 & 1.0  & -74.8 & 1.0 & 3.0 & 1.0 & - & - \\
 48913.412 & -43.0 & 1.0  & 105.6 & 1.0 & 19.0 & 1.0 & 50.9 & 1.0 \\
 48916.375 & -55.5 & 1.0  & 125.2 & 1.0 & - & - & - & - \\
 49246.472 & 76.6 & 1.0  & -67.1 & 1.0 & 6.6 & 0.4 & - & - \\
 49248.525 & -56.1 & 1.0  & 131.7 & 1.0 & 44.0 & 1.0 & 15.3 & 1.0 \\
 49250.537 & 101.6 & 1.0  & -93.6 & 1.0 & 58.9 & 0.4 & 1.7 & 1.0 \\
 49302.418 & -53.7 & 1.0  & 132.4 & 1.0 & 12.9 & 1.0 & 51.9 & 1.0 \\
 49307.374 & 91.3 & 1.0  & -87.2 & 1.0 & 23.0 & 0.4 & 34.6 & 0.4 \\
 49618.436 & 94.0 & 1.0  & -92.1 & 1.0 & - & - & - & - \\
 49622.425 & 60.6 & 1.0  & -41.1 & 1.0 & - & - & - & - \\
 49677.381 & -50.0 & 1.0  & 132.6 & 1.0 & - & - & - & - \\
 49971.535 & 93.8 & 1.0  & -78.7 & 1.0 & -9.8 & 1.0 & 53.5 & 0.4 \\
 49973.434 & -48.2 & 1.0  & 138.8 & 1.0 & - & - & - & - \\
 49973.535 & -51.1 & 1.0  & 142.9 & 1.0 & 17.0 & 0.4 & 26.0 & 0.4 \\
 50364.509 & 102.5 & 1.0  & -85.3 & 1.0 & 5.8 & 0.4 & 30.0 & 0.4 \\
 50366.371 & -48.5 & 1.0  & 146.0 & 1.0 & - & - & - & - \\
 50400.392 & 100.7 & 1.0  & -88.4 & 1.0 & 38.2 & 1.0 & -1.3 & 1.0 \\
 50404.270 & 91.1 & 1.0  & -66.5 & 1.0 & - & - & - & - \\
 50721.423 & 93.8 & 1.0  & -67.1 & 1.0 & -9.9 & 1.0 & 38.0 & 0.4 \\
 50721.501 & 96.1 & 1.0  & -77.1 & 1.0 & -11.3 & 1.0 & 40.2 & 1.0 \\
 50755.460 & -46.1 & 1.0  & 139.7 & 1.0 & 38.8 & 1.0 & -11.3 & 0.4 \\
 50757.384 & 104.0 & 1.0  & -86.7 & 1.0 & 34.0 & 1.0 & -2.7 & 1.0 \\
 50832.293 & 104.7 & 1.0  & -82.2 & 1.0 & -9.7 & 1.0 & 44.2 & 1.0 \\
 50833.156 & 57.7 & 1.0  & - & - & - & - & - & - \\
 51088.384 & -7.6 & 1.0  & 92.2 & 1.0 & - & - & - & - \\
 51089.481 & 106.0 & 1.0  & -85.2 & 1.0 & - & - & - & - \\
 51091.427 & -49.1 & 1.0  & 149.9 & 1.0 & -18.8 & 0.4 & 47.3 & 1.0 \\
 51093.343 & 103.8 & 1.0  & -79.4 & 1.0 & 3.6 & 0.4 & 22.0 & 0.4 \\
 51473.377 & -51.8 & 1.0  & 147.2 & 1.0 & 43.5 & 1.0 & -18.3 & 0.4 \\
 51475.281 & 107.5 & 1.0  & -85.7 & 1.0 & - & - & - & - \\
 51475.362 & 107.8 & 1.0  & -85.7 & 1.0 & 22.1 & 0.4 & 4.1 & 0.4 \\
 51803.489 & 95.8 & 1.0  & -66.7 & 1.0 & - & - & - & - \\
 51805.463 & -49.1 & 1.0  & 147.2 & 1.0 & 41.6 & 1.0 & -21.2 & 0.4 \\
 51807.463 & 107.0 & 1.0  & -86.9 & 1.0 & 5.1 & 0.4 & 22.7 & 0.4 \\
 52016.110 & -46.8 & 1.0  & 142.6 & 1.0 & - & - & - & - \\
 52180.462 & -49.0 & 1.0  & 146.2 & 1.0 & -14.7 & 1.0 & 43.8 & 1.0 \\
 52182.481 & 104.0 & 1.0  & -86.2 & 1.0 & 28.2 & 0.4 & 0.4 & 0.4 \\
 52327.104 & -48.8 & 1.0  & 145.5 & 1.0 & 35.9 & 1.0 & -0.9 & 0.4 \\
 52329.148 & 89.8 & 1.0  & -64.0 & 1.0 & -8.0 & 0.4 & 38.2 & 0.4 \\
 52537.494 & -50.0 & 1.0  & 137.7 & 1.0 & 35.0 & 1.0 & -4.0 & 0.4 \\
 52539.468 & 106.0 & 1.0  & -89.1 & 1.0 & 42.9 & 1.0 & -7.0 & 1.0 \\
 52541.378 & -46.3 & 1.0  & 144.8 & 1.0 & - & - & - & - \\
 52541.473 & -45.5 & 1.0  & 139.3 & 1.0 & 1.5 & 0.4 & 38.9 & 1.0 \\
 52705.198 & -45.8 & 1.0  & 126.6 & 1.0 & 48.5 & 1.0 & - & - \\
 52707.190 & 101.6 & 1.0  & -88.8 & 1.0 & - & - & - & - \\
 52709.127 & -54.6 & 1.0  & 144.6 & 1.0 & - & - & 51.1 & 1.0 \\
 52903.438 & 93.6 & 1.0  & -75.7 & 1.0 & - & - & - & - \\
 52941.345 & -50.8 & 1.0  & 141.1 & 1.0 & 51.1 & 1.0 & -6.8 & 0.4 \\
 53273.479 & -53.1 & 1.0  & 139.3 & 1.0 & 41.0 & 1.0 & 7.9 & 1.0 \\
 53278.490 & 93.6 & 1.0  & -84.9 & 1.0 & - & - & - & - \\
 53637.454 & -55.3 & 1.0  & 134.5 & 1.0 & 12.9 & 1.0 & 42.4 & 1.0 \\
 54001.454 & -42.5 & 1.0  & 113.7 & 1.0 & 3.6 & 0.4 & 66.6 & 1.0 \\
 54003.423 & 89.9 & 1.0  & -90.3 & 1.0 & - & - & - & - \\
 54005.351 & -58.5 & 1.0  & 137.5 & 1.0 & 62.1 & 1.0 & - & - \\
 54005.506 & -57.7 & 1.0  & 134.6 & 1.0 & 60.4 & 1.0 & 0.9 & 0.4 \\
\enddata
\tablenotetext{a}{An observation of Aa1 of unit weight has a standard error of 2.0 ${\rm km\,s^{-1}}.$}
\tablenotetext{b}{An observation of Aa2 of unit weight has a standard error of 2.3 ${\rm km\,s^{-1}}.$}
\tablenotetext{c}{An observation of Ab1 of unit weight has a standard error of 2.6 ${\rm km\,s^{-1}}.$}
\tablenotetext{d}{An observation of Ab2 of unit weight has a standard error of 2.6 ${\rm km\,s^{-1}}.$}
\end{deluxetable*} 

From 1984 January through 2006 September we obtained 82 spectrograms
of 88~Tau with the Kitt Peak National Observatory (KPNO) 0.9 m coud\'e 
feed telescope, coud\'e spectrograph, and a TI~CCD detector.  Sixty-eight 
spectrograms are centered in the red at 6430~\AA, cover a wavelength
range of about 80~\AA, and have a two pixel resolution of 0.21~\AA.  
Those spectra have signal-to-noise ratios of $\sim$250.  The remaining
14 spectrograms are centered in the blue at 4500~\AA, cover a 
wavelength range of 85~\AA, and have a resolution of 0.22~\AA. 
Signal-to-noise ratios of $\sim$300 are typical.  

\begin{deluxetable*}{lcccccccc}
\tablecaption{Fairborn Observatory Radial Velocity data for 88 Tau A\label{tab:datafairborn}}
\tablewidth{0pt}
\tablehead{
HJD-2400000.5 &  $V_{Aa1}$& Weight\tablenotemark{a} & $V_{Aa2}$ & Weight\tablenotemark{b} & $V_{Ab1}$ & Weight\tablenotemark{c} & $V_{Ab2}$ & Weight\tablenotemark{d} \\
                              &   (${\rm km\,s^{-1}}$) & & (${\rm km\,s^{-1}}$) & & (${\rm km\,s^{-1}}$) & & (${\rm km \,s^{-1}}$) & 
}
\startdata 
53020.204 & -38.5 & 1.0  & 118.0 & 1.0 & 50.8 & 1.0 & - & - \\
 53032.197 & 101.1 & 1.0  & -90.3 & 1.0 & -6.3 & 1.0 & 53.7 & 1.0 \\
 53052.273 & -43.0 & 1.0  & 126.0 & 1.0 & 50.9 & 1.0 & - & - \\
 53276.489 & -37.3 & 1.0  & 110.0 & 1.0 & - & - & 53.0 & 1.0 \\
 53285.468 & 85.9 & 1.0  & -69.9 & 1.0 & 10.8 & 1.0 & 41.8 & 1.0 \\
 53314.392 & 102.0 & 1.0  & -95.0 & 1.0 & 8.8 & 1.0 & 40.6 & 1.0 \\
 53350.390 & 93.6 & 1.0  & -85.3 & 1.0 & 46.3 & 1.0 & 7.7 & 1.0 \\
 53395.275 & -28.2 & 1.0  & 97.3 & 1.0 & -2.4 & 1.0 & 58.0 & 1.0 \\
 53405.184 & -44.4 & 1.0  & 127.0 & 1.0 & 39.9 & 1.0 & 11.9 & 1.0 \\
 53630.518 & -55.6 & 1.0  & 139.5 & 1.0 & 2.7 & 1.0 & 57.9 & 1.0 \\
 53644.477 & -48.0 & 1.0  & 127.2 & 1.0 & 38.3 & 1.0 & 20.1 & 1.0 \\
 53659.486 & -33.5 & 1.0  & 102.9 & 1.0 & 58.4 & 1.0 & - & - \\
 53700.387 & 90.5 & 1.0  & -82.3 & 1.0 & - & - & - & - \\
 53741.243 & -56.5 & 1.0  & 139.1 & 1.0 & -1.2 & 1.0 & 56.7 & 1.0 \\
 54191.132 & -61.0 & 1.0  & 136.8 & 1.0 & 7.3 & 1.0 & 63.7 & 1.0 \\
 54194.134 & -28.7 & 1.0  & 85.1 & 1.0 & - & - & - & - \\
 54194.156 & -30.2 & 1.0  & 89.5 & 1.0 & - & - & - & - \\
 54198.110 & -56.9 & 1.0  & 132.4 & 1.0 & - & - & - & - \\
\enddata
\tablenotetext{a}{An observation of Aa1 of unit weight has a standard error of 2.0 ${\rm km\,s^{-1}}.$}
\tablenotetext{b}{An observation of Aa2 of unit weight has a standard error of 2.3 ${\rm km\,s^{-1}}.$}
\tablenotetext{c}{An observation of Ab1 of unit weight has a standard error of 2.6 ${\rm km\,s^{-1}}.$}
\tablenotetext{d}{An observation of Ab2 of unit weight has a standard error of 2.6 ${\rm km\,s^{-1}}.$}
\end{deluxetable*} 

\begin{figure}
\figurenum{2}
\plotone{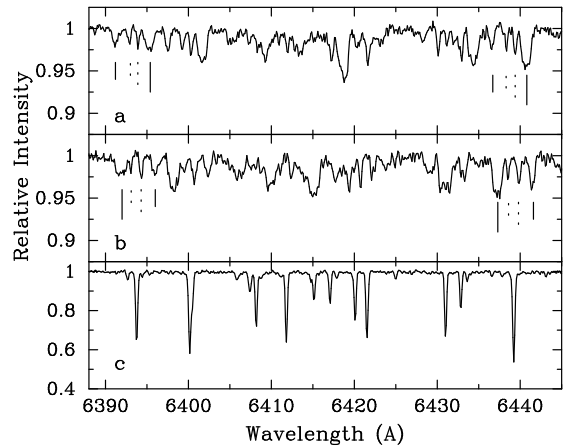}
\figcaption{Two spectra of 88 Tau, (a) JD 2,452,539.97 and 
(b) JD 2,454,006.01, compared with (c) the spectrum of the IAU radial 
velocity standard 10 Tau.  The four components of 88 Tau A are identified 
for two different lines. Solid tick marks indicate lines of the Aa (3.57 day 
period) binary with the longer tick mark identifying the primary.  Dotted tick 
marks indicate lines of the Ab (7.89 day period) binary with the longer tick mark 
identifying the primary. 
\label{fig2}}
\end{figure}

From 2004 January through 2007 April we acquired 29 spectrograms  
with the Tennessee State University 2 m automatic spectroscopic
telescope (AST), fiber-fed echelle spectrograph, and a 2048 x 4096 SITe
ST-002A CCD.  The echelle spectrograms have 21 orders, covering the
wavelength range 4920--7100~\AA\ with an average resolution of 0.17~\AA.
The typical signal-to-noise ratio is $\sim$50.
\citet{ew04} have given a more extensive description of the telescope,
situated at Fairborn Observatory near Washington Camp in the Patagonia
Mountains of southeastern Arizona, and its operation.

For the KPNO spectrograms we determined radial velocities with the
IRAF cross-correlation program FXCOR \citep{f93}, fitting  Gaussian
functions to the individual cross-correlation peaks.  
Double Gaussian fits were required to obtain individual velocities from blended cross-correlation 
peaks.  The IAU radial velocity standard
star 10~Tau was used as the cross-correlation reference star for the
red-wavelength spectrograms.  Its velocity of 27.9 km~s$^{-1}$ was adopted
from \citet{setal90}.  Lines in the wavelength region redward of 6445~\AA\
are not particularly suitable for measurement because most features are
blends, and there are a number of modest strength water vapor lines.
Thus, the radial velocities were determined from lines in the region
6385--6445~\AA.  However, this 60~\AA\ portion of the spectrum is
so small that a spectrum mismatch, caused by the varying strength of line
blends with temperature, between the A and F spectral type components of 
88~Tau and the F9~IV-V \citep{km89} spectral type of the reference star
10~Tau, can significantly alter the measured velocity.  Thus,
instead of cross-correlating this entire 60~\AA\ wavelength region, only
the wavelength regions around two or three of the strongest and 
least-blended lines, usually the Fe~I lines at 6394 and 6412~\AA\ 
plus the Ca~I line at 6439~\AA, were cross-correlated.

At blue wavelengths the Am star dominates the spectrum.  To compute 
velocities from those spectrograms, 68~Tau, spectral type A2~IV 
\citep{am95}, which has a velocity of 39.0 km~s$^{-1}$ \citep{f99}, 
was used as the reference star.  The region between 4485 and 4525~\AA\ 
was cross correlated. Velocities for our KPNO spectra
are given in Table \ref{tab:dataknpo}.

For the Fairborn Observatory AST spectra, lines in approximately 100 
regions, centered on the rest wavelengths \citep{metal66} of relatively 
strong lines (mostly of Fe~I and Fe~II) that were not strong blends,
were measured.  Lines at the ends of each echelle
order were excluded because of their lower signal-to-noise ratios.
A Gaussian function was fitted to the profile of each component.  
Double Gaussian fits were required to represent blended
components.  The difference between the observed wavelength and that 
given in the solar line list of \citet{metal66} was used to compute 
the radial velocity, and a heliocentric correction was applied.  The 
final mean velocity for each observation is given in Table~\ref{tab:datafairborn}.  
Unpublished velocities of several IAU standard stars with F dwarf 
spectral types indicate that the Fairborn Observatory velocities have 
a small zero-point offset of $-$0.3 km~s$^{-1}$ relative to the 
velocities of \citet{setal90}. 

\subsection{Preliminary spectroscopic analysis}
In a study of lithium in Am stars \citet{bc88} acquired two high-resolution
spectrograms of 88 Tau in the 6710~\AA\ region.  Comparing two
sets of lines in the two spectra, they reported detecting the lines of 
5 different components.  Figure~2 presents two spectra of 88~Tau~A in the 
6430~\AA\ region that show the two components of the 3.57 day binary near 
opposite nodes in their spectroscopic orbit, when the components have 
their maximum velocity separation.  Between the two ``outside'' 
lines are two additional weak components.  From a careful inspection of 
our KPNO spectra, as well as the ones obtained at Fairborn Observatory, 
we find lines of only 4 components rather than the 5 reported by 
\citet{bc88}.  In Figure~\ref{fig2} many lines of the Am star are $\sim$5\% deep, 
while the lines of the other 3 components typically have line depths 
$\leq$2.5\%.  Correctly identifying components in such a weak-lined 
and complex spectrum is not easy because in many spectra two or 
more of the components are blended. 

While the identification of the components of the 3.57 day binary is 
straightforward, the very weak lines of the other two components are 
similar in strength and line width, making it difficult to tell 
them apart.  To determine a preliminary orbital period, we initially 
examined only the latter portion of our KPNO velocities, obtained 
from MJD 50400 to 54000.  For each observation we computed the absolute
value of the velocity difference between the two components and then 
used those results as the input data for two different period finding 
approaches.  First, a sine curve was fitted to the velocity differences 
for trial periods between 1 and 100 days with a step size of 0.0005 
days.  The period with the smallest sum of the squared residuals was 
adopted as the best period.  Next, a search over a similar period 
range and with the same step size was done with the least string 
method \citep{betal70}.  Both searches resulted in a period of 3.9435 
days. Doubling this period produced an orbital period of 7.887 days.  
Separate analyses of our earlier KPNO velocities as well as the Fairborn 
Observatory velocities produced a similar orbital period.  We then 
adopted the 7.887 day period and computed a phase diagram to identify
correctly the components.  Afterward we compared those results with 
an attempt at visual identification, based on which set of 
lines appeared to be stronger in each spectrum.  The visual inspection
correctly identified the more massive component only about half of the 
time.  Apparently, the lines of these two components are similar 
enough that weak lines from other components and noise can significantly
affect the apparent line strengths.  Thus, we conclude that in our spectra it is not 
possible to differentiate the two components based on line strength.

\subsection{Orbital Models}

In modeling the hierarchical quadruple system we make the simplifying
assumption that the three orbital systems do not perturb each other during 
the time of our observations, i.e. we use three Keplerian orbital systems, one wide (Aa-Ab) and slow
(18-year period), and two short period systems: Aa1-Aa2, 3.57-day period,
and Ab1-Ab2, 7.89-day period.  Note that one cannot simply superimpose
the separation vectors from the three models; this is because the
PHASES observable is the angle between the two Centers-of-Light (COL)
of the short-period systems. 
\begin{eqnarray}\label{couplingEquation}
\overrightarrow{y_{\rm{obs}}} & = & \overrightarrow{r_{\rm{Aa-Ab}}} \nonumber \\
                                                     &     & + \frac{ R_{Aa} - L_{Aa}}{\left(1+R_{Aa}\right)\left(1+L_{Aa}\right)}\overrightarrow{r_{\rm{Aa1-Aa2}}} \nonumber \\
                                                     &     & - \frac{ R_{Ab} - L_{Ab}}{\left(1+R_{Ab}\right)\left(1+L_{Ab}\right)}\overrightarrow{r_{\rm{Ab1-Ab2}}}
\end{eqnarray}
Here $R_{Aa} = M_{\rm{Aa2}}/M_{\rm{Aa1}}$ is the Aa component 
mass ratio and $L_{Aa} = L_{\rm{Aa2}}/L_{\rm{Aa1}}$ the luminosity ratio,
while $R_{Ab} = M_{\rm{Ab2}}/M_{\rm{Ab1}}$ and $L_{Ab} = L_{\rm{Ab2}}/L_{\rm{Ab1}}$
are the corresponding ratios for the Ba--Bb sub-system. 
Including this coupling term for astrometric data is 
important when a full analysis, including radial velocity data, is made.

\section{Results}

The best-fit orbital model was found with an iterative non-linear
least-squares minimization scheme. The best-fit parameters 
are found in Table \ref{tab:fit}. The reduced $\chi_r^2$ of the combined fit
to PHASES, radial velocity, and previous differential astrometry data
is 1.37.  This combined set has 378 data points (49 of which are 
two-dimensional astrometric points) and  23 free parameters. 
The fits to the astrometric and radial-velocity data for the various 
subsystems are shown in Figures 3--7.
We find that the two short-period systems have eccentricities 
consistent with zero, and we therefore held these parameters 
fixed at zero for the fit. The time of maximum primary apparent velocity 
is chosen as zero orbital phase. 
To investigate the consistency of two astrometric data sets we
also re-ran the fit, without including the previous astrometry. We find that 
the final results are consistent to within one sigma; including the 
previous astrometry does reduce certain parameter uncertainties by a small amount. 
In Table \ref{tab:fit}  we compare our best-fit values with those available in the 
literature \citep{abt,balega99} where possible; we find generally good 
agreement.

Finally, we considered the possibility of an additional massive body in the
system.  At this point, a full mass-period phase space search for a fifth
component is computationally prohibitive and beyond the scope of this
investigation.  However, through the course of the investigation, we discovered
a possible periodicity that warranted follow-up.  Before we had determined
the 7.89 day period of the Ab1-Ab2 system, we had searched the
PHASES data for this third orbit signal with the
period-searching program used to find the signatures of additional
companions\citep{limits}; because the Aa-Ab and Aa1-Aa2 periods
were known already, these orbits were seeded at their best fit double Kepler values. 
While radial velocities eventually isolated the period for Ab1-Ab2 as 7.89 days, this
astrometric search additionally identified a curious improvement near 56 days.  The
size of that orbit could have corresponded either to a brown dwarf or to a nearly equal
luminosity star-star subsystem.  Inspired by this identification, a 5-component, quadruple
Keplerian model was seeded with the three known orbits (Aa-Ab, Aa1-Aa2, and
Ab1-Ab2) and a fourth perturbation with a period of 56 days. Separately, the harmonic of
28 days was also investigated.  This model included all astrometric and radial velocity data. 
The 5-component model with a candidate period near 28 days (C/Ab mass ratio $\sim$ 0.044) 
yields a significant improvement in the reduced $\chi^2$ of the fit over a 4-body
model  (491 for 30 parameters and 397 degrees of freedom vs. 547 for 23 parameters and
404 degrees of freedom, corresponding to a $\sim 2.3\sigma$ improvement from the
$\chi^2$ ratio test).   The candidate period near 56 days yields a best-fit $\chi^2 \sim 474$
for a $3.7\sigma$ improvement over the 4-component model, yet with a very large best-fit eccentricity (0.93). 
However, we note that the 28-day period is close to the product of the
Aa and Ab orbital periods ($\sim28.1$ days) and the 56 day period is a harmonic of this,
raising suspicion regarding the reliability of the solution.  Hence the astrophysical
significance of this additional component remains unclear.  We do not claim this as a
detection, but it is an intriguing result worthy of observational follow-up.

\begin{figure*}[t]
\figurenum{3}
\plottwo{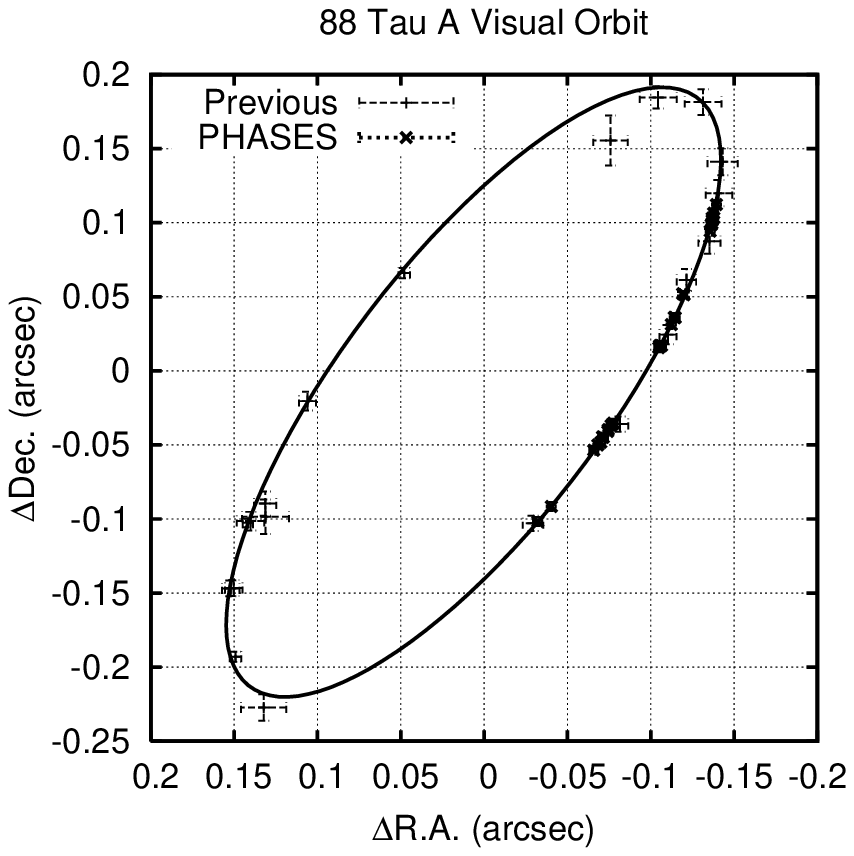}{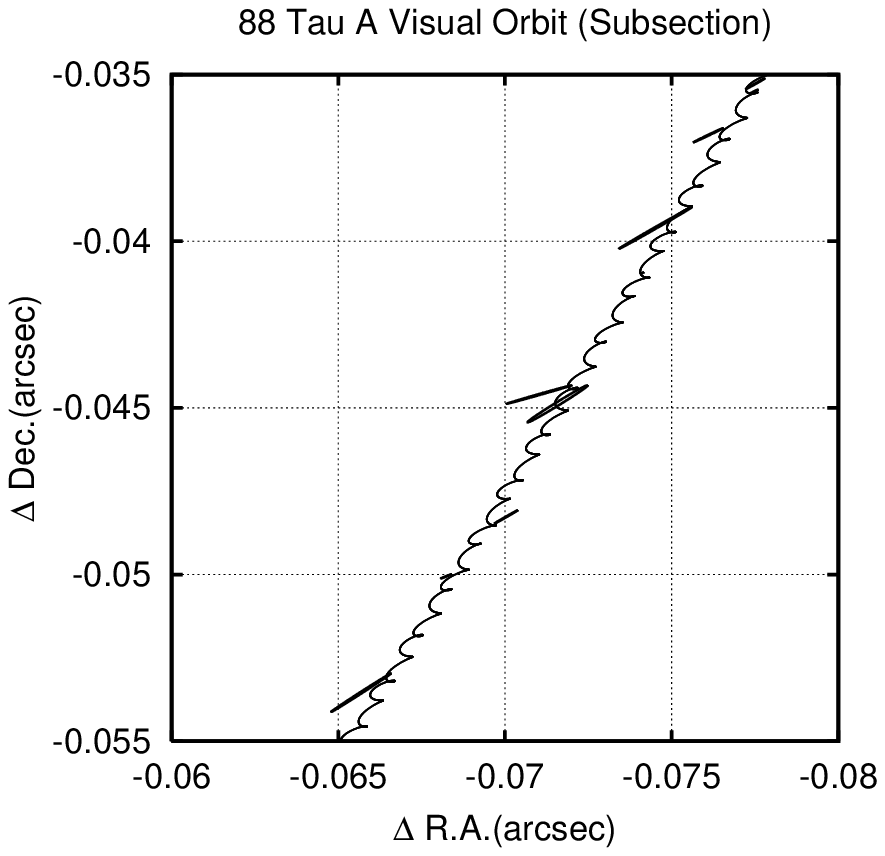}
\caption[]{\label{fig:astro_wide} (left)The best-fit visual orbit of the 88 Tau Aa-Ab system, together
with previously available astrometric data, and our PHASES astrometry. Note that the 
error ellipses of the PHASES data appear smaller than the points used to indicate the 
data. (right) A close-in view 
of a subsection of the PHASES astrometry, together with the best-fit orbital model.}
\end{figure*}

\begin{figure*}[t]
\figurenum{4}
\plottwo{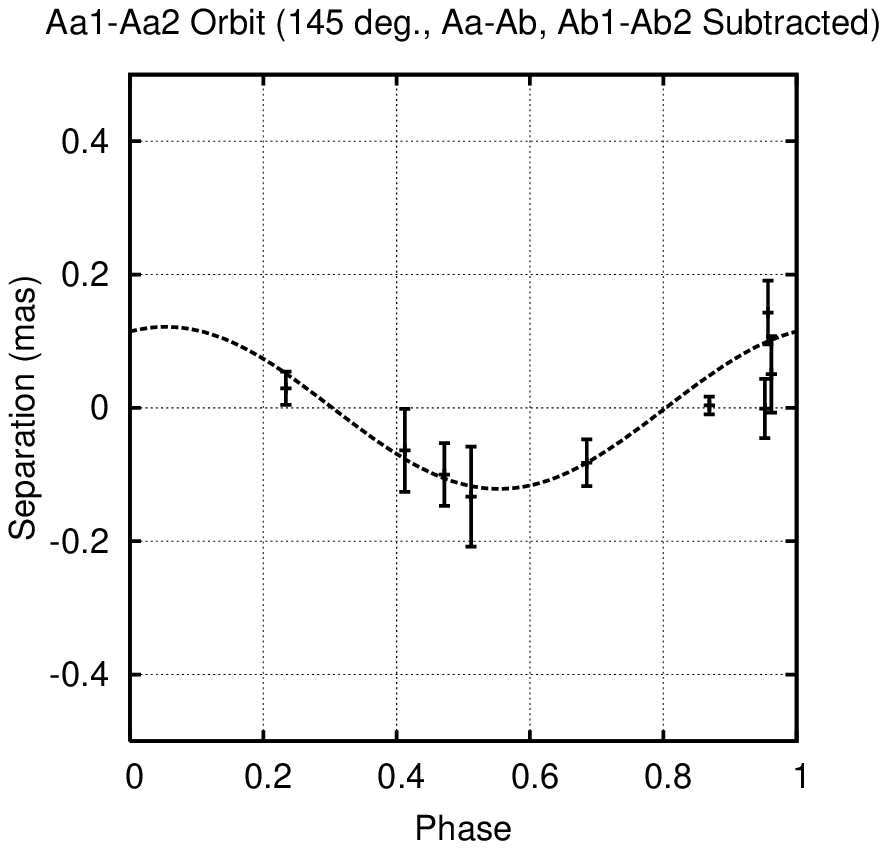}{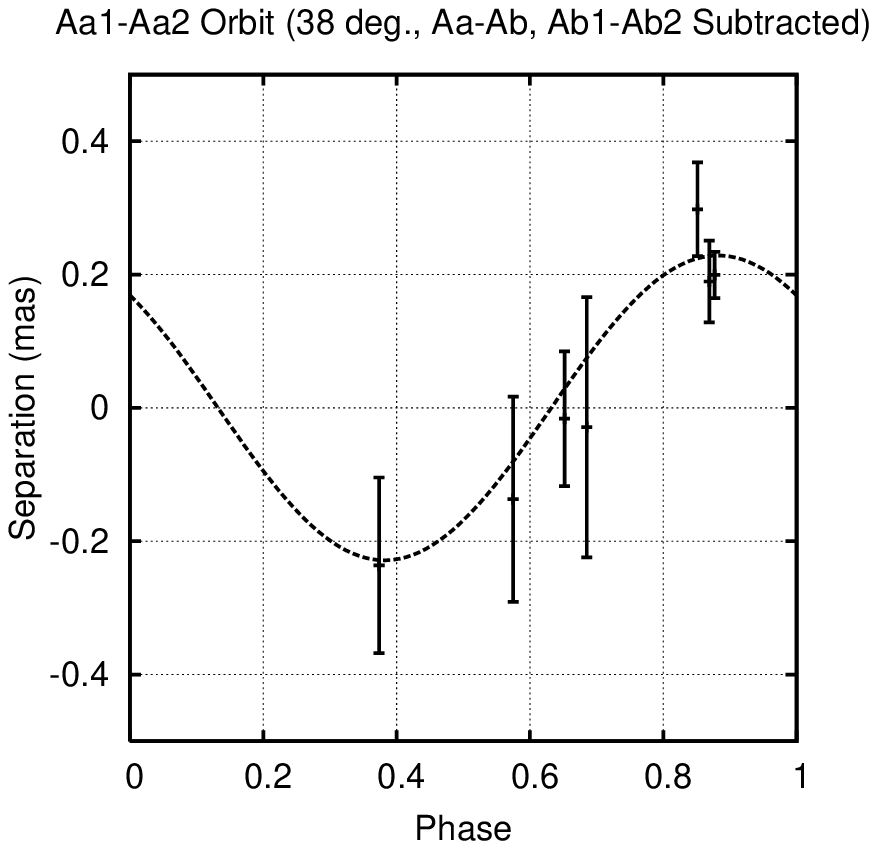}
\caption[]{\label{fig:astro_narrow_01} The astrometric orbit of the 88
  Tau Aa1-Aa2 subsystem, projected along two different 
  axes (145 degrees East of North deg on the left, 38 degrees on the right.) In each case the
  motion of the Aa--Ab and other sub-system has been removed. The
  axes correspond to the two most common orientations of the minor axis
  of the positional error ellipses (which vary slightly from night to
  night, and between baselines).  For clarity, only those observations where the projected
  uncertainty is less than 300 $\mu$as have been included in the plot
  (all observations are included in the fit.)  }
\end{figure*}

\begin{figure}[t]
\figurenum{5}
\plotone{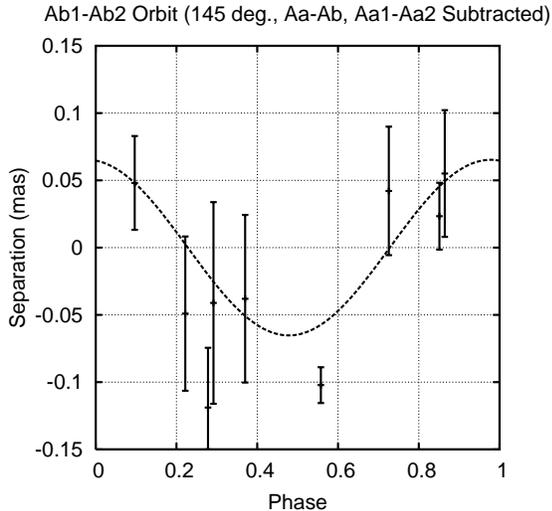}
\caption[]{\label{fig:astro_narrow_02} The astrometric orbit of the 88
  Tau Ab1-Ab2 sub-system.  The
  motion of the Aa--Ab and other sub-system has been removed. The
  separations shown are projected along an axis oriented 145 degrees
  East of North, corresponding to mean orientation of the minor axis
  of the positional error ellipses (which vary slight from night to
  night).  For clarity, only those observations where the projected
  uncertainty is less than 300 $\mu$as have been included in the plot
  (all observations are included in the fit).  }
\end{figure}

\begin{figure*}[t]
\figurenum{6}
\plottwo{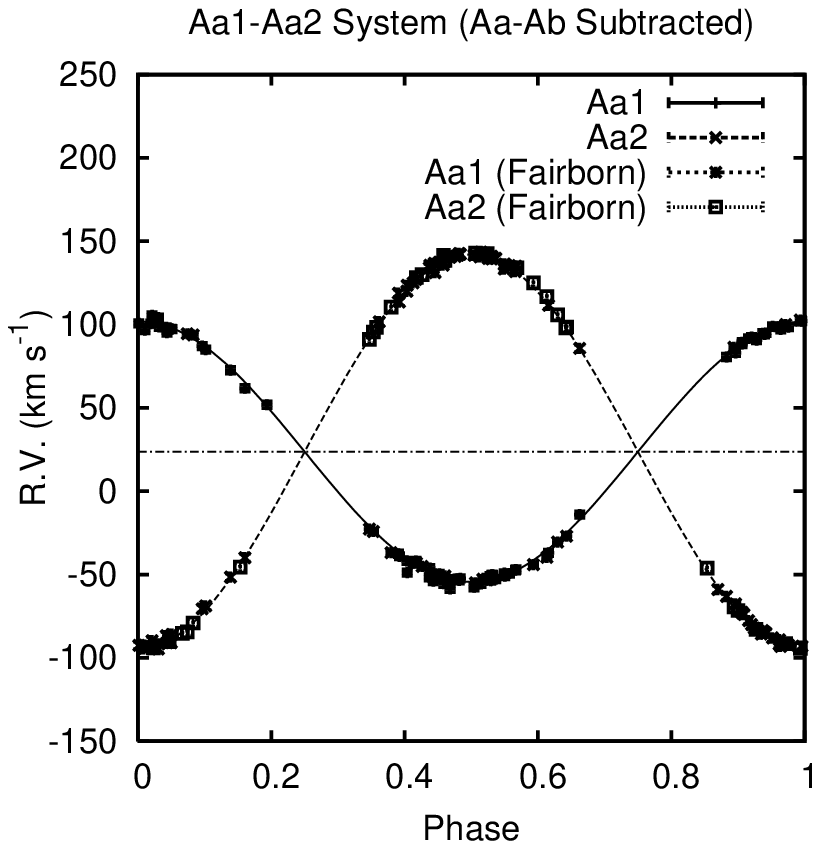}{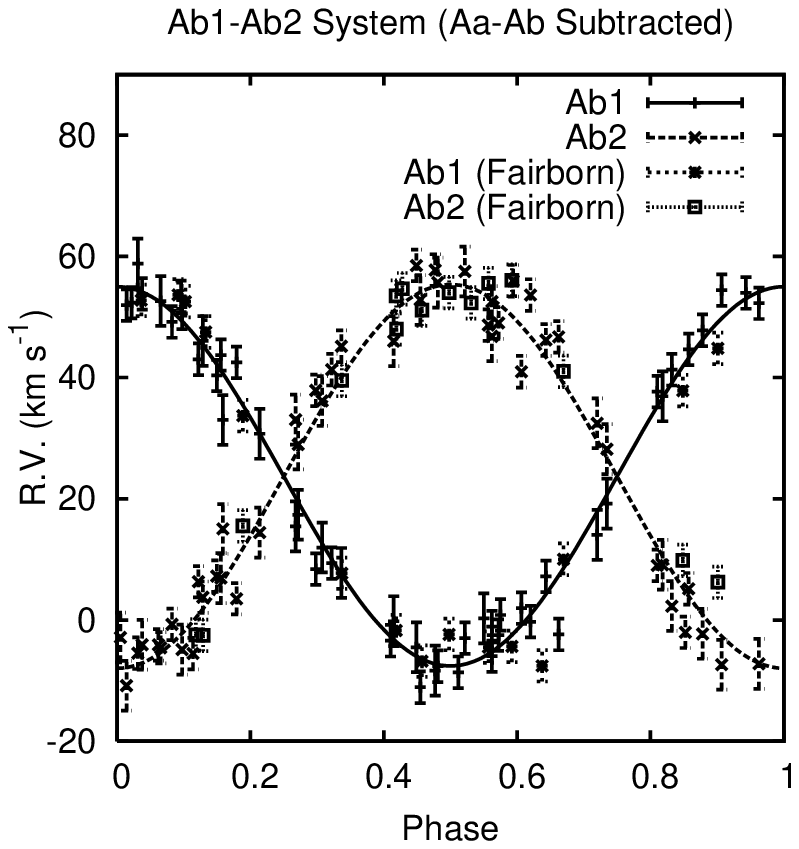}
\caption[]{\label{fig:rv_narrow} (left) The measured and model radial velocities of the 
Aa1--Aa2 subsystem, phased about the best-fit orbital model, and with the 
motions due to the Aa--Ab orbit removed. (right) Measured and model 
radial velocities of the Ab1--Ab2 subsystem, with the Aa- -Ab motion removed. }
\end{figure*}

\begin{figure}[t]
\figurenum{7}
\plotone{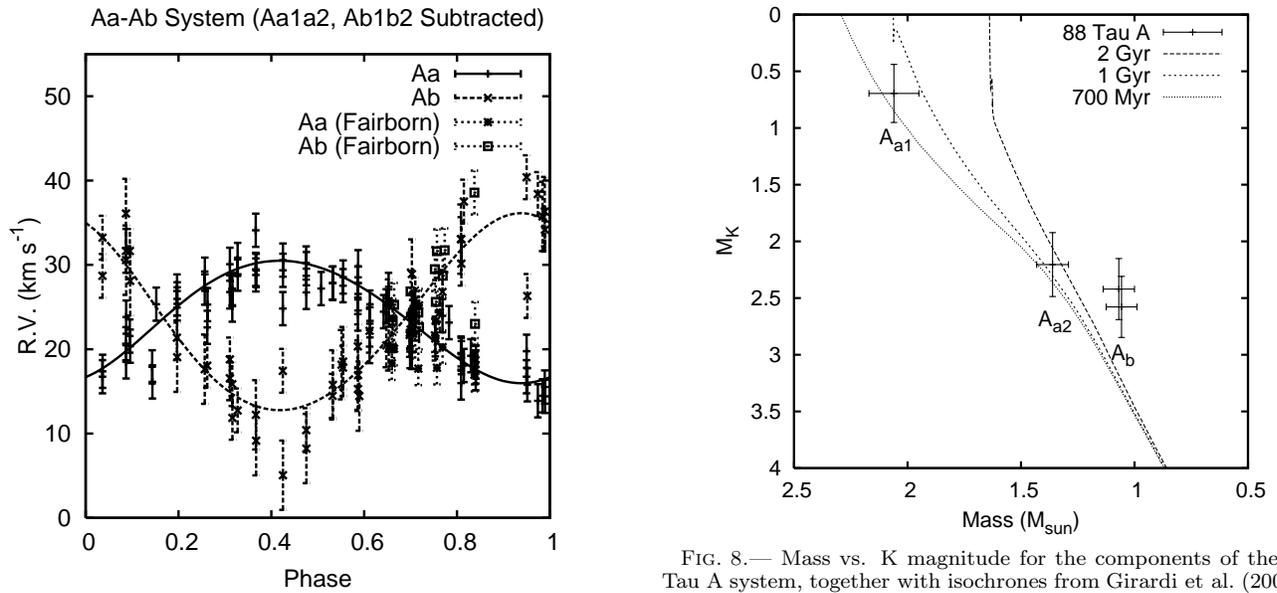}
\caption[]{\label{fig:rv_wide} The measured and modeled radial velocities 
 of the 88 Tau Aa-Ab system, with the motion due to the Aa1--Aa2 and Ab1--Ab2
 systems subtracted. }
\end{figure}

\begin{deluxetable}{l ll}
\tablecaption{Best-fit Orbital Parameters for 88 Tau A\label{tab:fit}}
\tablewidth{0pt}
\tablehead{
 Parameter & Value \& Uncertainty & Previous Value 
}
\startdata 
$ \chi^2     $ & $   547   $ &  \\
$ \chi_{r}^2   $ & $     1.37 $ & $  $ \\
$ {\rm No. Param.} $ & $    23 $ & $  $ \\
$ {\rm P_{AaAb}(days)} $ & $  6585  \pm 12 $ &  $6593\pm44\tablenotemark{c}$ \\
$ {\rm e_{AaAb}} $ & $     0.0715\pm 0.0026 $ & $0.083\pm0.008\tablenotemark{c}$ \\
$ {\rm i_{AaAb}(deg.)} $ & $    69.923 \pm 0.048 $ & $70.4\pm0.4\tablenotemark{c}$  \\
$ {\rm \omega_{AaAb}(deg.)} $ & $   205.7  \pm 1.2 $ & $222.0\pm3.0\tablenotemark{c}$\\
$ {\rm T_{AaAb} (HMJD)} $ & $ 55261  \pm 22 $ & \\
$ {\rm \Omega_{AaAb}(deg.)} $ & $   146.734 \pm 0.067 $ & $146.6\pm0.3\tablenotemark{c}$\\
$ {\rm M_{Aa}(M_{\odot})} $ & $     3.42 \pm 0.18 $ & \\
$ {\rm M_{Ab}(M_{\odot})} $ & $     2.13  \pm 0.13 $ & \\
$ {\rm d (pc)} $ & $    50.70  \pm 0.88 $ & \\
\tableline

$ {\rm P_{Aa1Aa2}(days)} $ & $     3.571096 \pm 0.000003 $ &  $3.571391\pm0.000008\tablenotemark{d}$\\
$ {\rm e_{Aa1Aa2}} $ & $     0.0   $ & $0\tablenotemark{d}$\\
$ {\rm i_{Aa1Aa2}(deg.)} $ & $   110.6  \pm 2.7 $ & \\
$ {\rm \omega_{Aa1Aa2}(deg.)} $ & $     0.0  $ & \\
$ {\rm T_{0,Aa1Aa2}(MHJD)} $ & $ 53389.3824  \pm 0.0030 $ & $43108.103\pm0.013\tablenotemark{d}$\\
$ {\rm \Omega_{Aa1Aa2}(deg.)} $\tablenotemark{a} &$  287.5 \pm 1.8 $ & \\
$ {\rm M_{Aa2}/M_{Aa1}} $ & $     0.6602  \pm 0.0028 $ & \\
$ {\rm L_{Aa2}/L_{Aa1}}$(K-band)\tablenotemark{a}  & $     0.249 \pm 0.035 $ & \\
\tableline

$ {\rm P_{Ab1Ab2}(days)} $ & $     7.886969 \pm 0.000066 $ & \\
$ {\rm e_{Ab1Ab2}} $ & $     0.0 $ &  \\
$ {\rm i_{Ab1Ab2}(deg.)} $ & $    27.23  \pm 0.72 $ & \\
$ {\rm \omega_{Ab1Ab2}(deg.)} $ & $     0.0  $ &  \\
$ {\rm T_{0,Ab1Ab2}(MHJD)} $ & $ 52507.31  \pm 0.02 $ & \\
$ {\rm \Omega_{Ab1Ab2}(deg.)} $\tablenotemark{b} & $    34.0  \pm 8.2 $ & \\
$ {\rm M_{Ab2}/M_{Ab1}} $ & $     0.988 \pm 0.024 $ & \\
$ {\rm L_{Ab2}/L_{Ab1}}$(K-band)\tablenotemark{b}  & $     0.865 \pm 0.028 $ & \\
$ {\rm V_0 (KPNO, km\,s^{-1})}  $ & $ 23.70  \pm 0.17 $ & \\
$ {\rm V_0 (Fairborn,km\,s^{-1})} $ & $ 23.91 \pm 0.31 $ & \\
\enddata
\tablenotetext{a}{An alternate, but disfavored, solution has $L_{Aa2}/L_{Aa1} = 1.48 $ and $ \Omega_{Aa1Aa2} =326$.}
\tablenotetext{b}{An alternate solution has $L_{Ab2}/L_{Ab1} = 1.10$ and $\Omega_{Ab1Ab2} = 205$.}
\tablenotetext{c}{From \citet{balega99}}
\tablenotetext{d}{From \citet{abt}}
\end{deluxetable} 

\begin{deluxetable}{l ll}

\tablecaption{Derived System Parameters for 88 Tau A\label{tab:params}}
\tablewidth{0pt}
\tablehead{
 Parameter & Value & Uncertainty
}
\startdata 
$ {\rm \Phi_{AaAb-Aa1Aa2}(deg.)} $ &  143.3 & $ \pm 2.5 $ \\
$ {\rm \Phi_{AaAb-Ab1Ab2}(deg.)} $ &  82.0\tablenotemark{a} & $ \pm 3.3 $ \\
$ {\rm \Phi_{Aa1Aa2-Ab1Ab2}(deg.)} $ &  115.8\tablenotemark{b} & $ \pm 4.6 $ \\
$ \pi {\rm (asec)} $ &  0.01973 & $ \pm 0.00034 $ \\
$ {\rm a_{AaAb}(milliarcsec)} $ &  240.1 & $ \pm  5.3 $ \\
$ {\rm a_{Aa1Aa2}(milliarcsec)} $ &  1.359 & $ \pm  0.034 $ \\
$ {\rm a_{Aa1Aa2,col} (milliarcsec)} $ &  0.270 & $ \pm  0.032  $ \\
$ {\rm a_{Ab1Ab2}(milliarcsec)} $ & 1.967 & $ \pm  0.054 $ \\
$ {\rm a_{Ab1Ab2,col}(milliarcsec)} $ & $0.065 $ & $ \pm  0.020 $ \\
$ {\rm a_{AB}(AU)} $ &  12.17 & $ \pm  0.17$ \\
$ {\rm a_{Aa1Aa2}(AU)} $ & 0.0689 & $ \pm  0.0012 $ \\
$ {\rm a_{Ab1Ab2}(AU)} $ & 0.0997 & $ \pm  0.0021 $ \\
$ {\rm M_{Aa1} (M_{\odot})} $ & 2.06& $ \pm  0.11$ \\
$ {\rm M_{Aa2} (M_{\odot})} $ & 1.361 & $ \pm  0.073$ \\
$ {\rm M_{Ab1} (M_{\odot})} $ & 1.069 & $ \pm  0.069$ \\
$ {\rm M_{Ab2} (M_{\odot})} $ & 1.057 & $ \pm  0.068$ \\
$ M_{K,Aa1} $ & 0.69    & $ \pm  0.26$ \\
$ M_{K,Aa2}  $ & 2.20    & $ \pm  0.28$ \\
$ M_{K,Ab1}  $ & 2.31 & $ \pm  0.27$ \\
$ M_{K,Ab2}  $ & 2.00 & $ \pm  0.27$ \\
\enddata
\tablenotetext{a}{An alternate solution has $\Phi_{AaAb-Ab1Ab2}=58$ deg. if $L_{Ab2}/L_{Ab1} >1$.}
\tablenotetext{b}{An alternate solution has $\Phi_{Aa1Aa2-Ab1Ab2}=107$ deg. if $L_{Ab2}/L_{Ab1} > 1$.}
\tablecomments{The parameters derived from the best-fit model values in Table \ref{tab:fit}
and their uncertainties. ``COL" refers to the amplitude of the motion of the Center of Light of the 
subsystem in question. Note that our combined astrometry and radial velocity 
model fits for the system masses directly.}
\end{deluxetable} 

\subsection{Relative Orbital Inclinations}

The mutual inclination $\Phi$ of two orbits is given by
\begin{equation}\label{MutualInclination}
\cos \Phi = \cos i_1 \cos i_2  + \sin i_1 \sin i_2 \cos\left(\Omega_1 - \Omega_2\right)
\end{equation}
\noindent where $i_1$ and $i_2$ are the orbital inclinations and $\Omega_1$ and $\Omega_2$ are the 
longitudes of the ascending nodes. For this quadruple system we derive 
three separate mutual inclinations, corresponding to the three possible
pairwise comparisons of the three orbits in this system (i.e. $Aab-Aa1Aa2$, $Aab-Ab1Ab2$ and 
$Aa1Aa2-Ab1Ab2$). The resulting values found from our combined 
orbital solutions are given in Table \ref{tab:params}.  The large 
mutual inclinations indicate that the system orbits are not co-planar.

It should be noted that even with both COL-astrometry and radial-velocity data, 
there exists a parameter degeneracy corresponding to an exchange of the 
ascending and descending nodes together with a change in the luminosity 
ratio (interchanging which is the brighter star). Given one solution for the 
mass and luminosty ratios ($R$ and $L_1$), the other possible luminosity 
ratio can be found from
\begin{equation}
L_2 = \frac{2R+RL_1 - L_1}{1+2L_1-R}
\end{equation}
In a quadruple stellar system such as 88 Tau A there are 4 possible 
model solutions. However, as can be seen in Fig. \ref{fig2}
the luminosity of the $Aa1$ component is clearly greater than the $Aa2$
component, hence we                                                                                                                                                                                                                                                                                                                                                                                                                                                                                                                                                                                                                                                                                                                                              choose the solution where $L_{Aa1Aa2} = 0.24$.
However, given the nearly-equal masses of the $Ab1$ and $Ab2$
components, it is not entirely clear which is the preferred solution
($L_{Ab1Ab2} = 0.87$ or $1.10$), and we calculate the two 
possible values for the corresponding mutual inclinations. 

The two possible mutual inclination values of the Ab system 
are both in the range for inclination-eccentricity 
oscillations of 39.2 -- 140.8 degrees \citep{kozai}, while the Aa 
system is not (though only 1 standard deviation away from the limit).  
During these``Kozai-cycles" the orbital eccentricity varies
on a timescale $\propto P_{out}^2/P_{in} (\sim 10^4$ years for
the $Ab$ system;  Kiseleva-Eggleton \& Eggleton 2001\nocite{kis01}),
and in the absence of damping factors the eccentricity 
values would range from $\sim 0$ to $\sim 0.98$ (Kiseleva et al., 1998\nocite{kis98};
however, see Kiseleva-Eggleton \& Eggleton 2001 for a discussion 
of various factors that may limit these excursions).
A full treatment of the dynamics of this system, including the 
effects of tidal friction, quadrupolar distortion and general relativity 
is beyond the scope of this paper.

Recently, \citet{ft07} studied the effect that Kozai oscillations 
would have on the distributions of orbital properties 
of triple systems. Systems with mutual inclinations in the 
range where Kozai cycles occur evolve rapidly as tidal 
dissipation during the high-eccentricity (and hence close approach)
phase of the oscillation causes the orbit of the inner binary to shrink. 
The resulting mutual  inclination distribution is strongly 
peaked near the critical values of 39 and 141 degrees. 
It is therefore interesting to note that the Aa system is 
so close to one of these predicted values.  The Ab system
is more ambiguous: the preferred (albeit only weakly) solution 
for the Ab system yields a mutual inclination that would
be expected to result in rapid orbital evolution. Interestingly, the
second possible solution is also relatively close to the 
critical limit. Clearly, further observations will be needed 
in order to remove the ambiguity in the node and hence mutual 
inclination.  

\subsection{Component Masses and Distance}

This study represents the first determination of the orbital inclinations
in this system, and hence the first time the masses have been determined;
the precision achieved is $\sim$ 5\% for the Aa components 
and $\sim 6$ \% for the Ab components. The parallax is found to
be $ 19.73  \pm 0.34$ mas (1.7\% uncertainty), placing the 
system at a greater distance than that estimated by {\em Hipparcos} ($21.68 \pm 0.82$ mas; ESA 1997\nocite{hipparcos}).
Our greater distance resolves the mass/luminosity 
discrepancy pointed out by \citet{balega99}, which arises if one 
assumes the {\em Hipparcos} distance to this system.

\citet{stal98} and \citet{balega99} stated
that the 3.57 day binary is eclipsing but gave no reference for
this claim.  Our inclination for that binary is $110.6\arcdeg$, a value
that is about $10\arcdeg$ above the maximum value required for 
eclipses to occur if canonical values are assumed for the radii of 
the two stars \citep{g92}.  Thus, the system is not eclipsing.  
We also note that because of our lower inclination, our masses are 
about $0.2 M_{\odot}$ larger than those adopted by \citet{balega99}.

\subsection{Component Luminosities}

As part of the combined astrometric and radial velocity fit we can 
solve for the K-band luminosity ratios of the components; this 
is because the distance and subsystem total masses are
essentially determined by the observations of the wide Aa--Ab system,
while the subsystem mass ratios are found from the subsystem
radial velocities; this leaves only the component luminosity ratios 
dependent on the size of the observed astrometric perturbation. 

\begin{figure}[t]
\figurenum{8}
\plotone{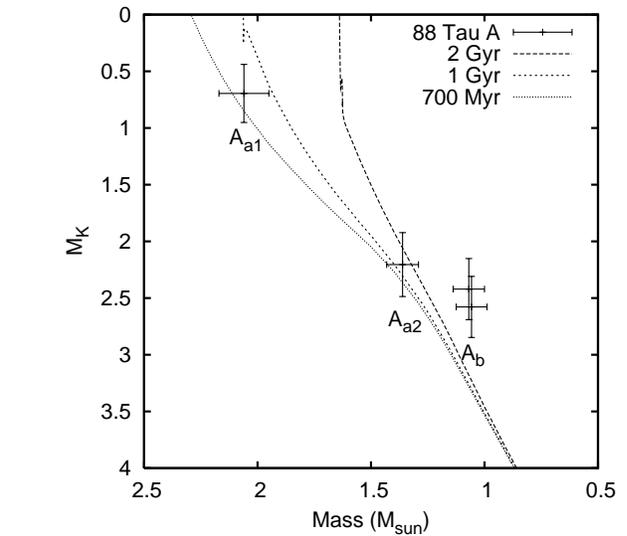}
\caption[]{\label{fig:massk} Mass vs. K magnitude for the components 
of the 88 Tau A system, together with isochrones from \citet{g02}.}
\end{figure}

While PTI cannot provide precise determinations of the total
system magnitude $m_K$ or the Aa--Ab system differential 
magnitude $\Delta m_K$, these can be found in the literature.
\cite{balega01} gave $\Delta m_{K'} = 1.29 \pm 0.12$ for the 
Aa-Ab system.  We derive the resulting absolute K magnitudes 
and list them in Table \ref{tab:params}; the results are dominated by
the uncertainty in the total magnitude. Note that we assume the 
solution where $L_{Ab2}/L_{Ab1} <1$.
We compare the determined masses and K-band absolute 
magnitudes to published theoretical 
isochrones \citep{g02} in the range 0.7--2 Gyr (Z=0.019). We find that the 
Aa components are consistent with the isochrones, but the
Ab system components appear over-luminous by $\sim$0.5 mag
(Figure \ref{fig:massk}).  However, this must be considered a preliminary finding
since it is dependent on a single measurement of the Aa--Ab 
magnitude difference, and we have not taken into account the 
(presumably minor) effects of the slightly different bandpasses ($K$ vs. $K'$).

\subsection{Spectral Classes and $v$ sin$i$ }

\citet{sf90} identified several luminosity-sensitive and temperature-sensitive
line ratios in the 6430-6465~\AA\ region.  Those critical line ratios and
the general appearance of the spectrum were employed as spectral-type
criteria.  However, for stars that are hotter than about early-G, the line
ratios in the 6430~\AA\ region have little sensitivity to luminosity, so
only the spectral class of an A or F star can be determined.  The luminosity
class is found by computing the absolute visual magnitude with the
{\it Hipparcos} or our orbital parallax and comparing that magnitude to
evolutionary tracks or a table of canonical values for giants and dwarfs.

The red wavelength spectrum of 88 Tau was compared with those of 
slowly rotating Am, F and G dwarfs. These reference stars, identified 
mostly from the lists of \citet{am95}, \citet{km89}, and \citet{f97}, 
were observed at KPNO with the same telescope, spectrograph, and 
detector as our spectra of 88 Tau.  With a computer program developed 
by \citet{hb84} and \citet{b85}, various combinations of reference-star 
spectra were rotationally broadened, shifted in radial velocity, 
appropriately weighted, and added together in an attempt to reproduce 
the spectrum of 88~Tau in the 6430~\AA\ region.  
\citet{am95} classified 88 Tau as an Am star with spectral classes of
A4, A6, and A7 for the calcium, hydrogen, and metal lines, respectively.
Their classification of HR~3526 was identical to that of 88~Tau, so
we adopted the spectrum of HR~3526 as the proxy for the primary of
the 3.57 day binary, which dominates the spectrum at blue wavelengths
and is still the strongest component in our red-wavelength region (Fig.~2).  
The 3 subclass difference between the calcium K line type and metal line 
type indicates that this star is a marginal or mild Am star \citep{ab69}.
A good fit to the lines of the 3.57 day secondary was produced by Procyon, 
spectral type F5~IV-V \citep{jm53}.  Lines of the two components in the 
7.89 day binary are similar in strength and rotation and were well 
represented by a spectrum of 70 Vir, spectral type G4~V \citep{km89}.
A fit with HR 483, spectral type G1.5 V \citep{km89}, used 
as a proxy for the 7.89 day binary pair, was nearly as good.
Thus, the spectral classes of the four stars are A6m, F5, G2-3:, and G2-3:,
where the colon indicates that the spectral class is more uncertain than
usual because of the weakness of the lines. 
As shown by their positions in Figure~8, all four stars are dwarfs.
The abundances of Procyon, 70 Vir, and HR 483 are essentially 
solar, indicating that the abundances of the components of 88~Tau, except for the Am star, 
are also close to solar.   

The continuum intensity ratio of our best reference star combination, 
fitted to the spectrum of 88~Tau, is 0.79:0.11:0.05:0.05.
If we adopt the continuum intensity ratios as the luminosity
ratios at 6430~\AA, we obtain a magnitude difference of 2.1 $\pm$ 0.3 
between the 3.57 day pair, and 2.4 $\pm$ 0.3 between the astrometric 
components, where the uncertainties are estimated. The 6430~\AA\ 
wavelength is about 0.6 of the way between the center of the Johnson 
$V$ and $R$ bandpasses.

With the procedure of \citet{f97}, we determined projected rotational
velocities for the 4 components of 88~Tau from 10 KPNO red-wavelength
spectra.  For each spectrum the full-widths at half-maximum of 2 or 3 
unblended lines in the 6430~\AA\ region were measured and the results 
averaged for each component.  The instrumental broadening
was removed, and the calibration polynomial of \citet{f97} was used to
convert the resulting broadening in angstroms into a total line broadening
in km~s$^{-1}$.  From \citet{f97,f03} we assumed a 
macroturbulence of 0.0 for the Am star, 4~km~s$^{-1}$ for the mid-F star, 
and 3 km~s$^{-1}$ for the G stars.  The resulting {\it v}~sin~{\it i} 
values are 37 $\pm$ 2 and 17 $\pm$ 2 km~s$^{-1}$ for the primary and 
secondary of the 3.57 day binary, respectively, and 5 $\pm$ 3 km~s$^{-1}$ 
for both components of the 7.89 day binary.  Our value for the Aa1 
component is consistent with the determination of {\it v}~sin~{\it i} =
36 km~$s^{-1}$ by \citet{r02}.

To determine whether the rotational velocities of the binary components
are synchronized, we assumed that the rotational axis of each component 
is parallel to its orbital axis.  We then computed the equatorial 
rotational velocity for each component and compared it with its expected 
synchronous velocity, computed with its canonical radius \citep{g92}. 
The resulting equatorial rotational velocities are 40, 18, 11, and 11 
km~s$^{-1}$, for components Aa1, Aa2, Ab1, and Ab2, respectively.  The 
computed synchronous rotational velocities are 24, 19, 6.5, and 6.5 
km~s$^{-1}$, respectively.
 
Component Aa1, the Am star and primary of the 3.57 day binary, is 
the only one of the four components that does not have a convective
atmosphere.  It is also the only component that is clearly not
rotating synchronously: its equatorial rotational velocity is
1.67 times faster than synchronous.  On the other hand its F5 
companion, component Aa2 is rotating synchronously.  Because of 
the relatively large uncertainties of our v sin i measurements for 
Ab1 and Ab2, it is possible that the two components of the 7.89 day 
are also synchronously rotating.

\section{Conclusion}

PHASES interferometric astrometry has been used together with 
radial velocity data to measure the orbital
parameters of the quadruple star system 88 Tau A, and in particular
to resolve the apparent orbital motion of the close Aa1-Aa2 and Ab1-Ab2 pairs.
We have made the first determination of the period of the 
Ab binary system and found it to consist of a pair of nearly equal-mass
G stars. The amplitude of the Ab1-Ab2 Center-of-Light motion is only $\sim 65 \mu$as, 
indicating the level of astrometric precision attainable
with interferometric astrometry. We are able to resolve the orbital
motion of all of the components, and hence determine the 
orbital inclinations and component masses with a precision of 
a few percent. Finally, we are able to determine the mutual 
inclinations of the various orbits. 

\acknowledgements We wish to acknowledge the extraordinary
observational efforts of K. Rykoski. Observations with PTI are made
possible thanks to the efforts of the PTI Collaboration, which we
acknowledge. This research has made use of services from
the Michelson Science Center, California Institute of Technology,
http://msc.caltech.edu.  Part of the work described in this paper was
performed at the Jet Propulsion Laboratory under contract with the
National Aeronautics and Space Administration. This research has made
use of the Simbad database, operated at CDS, Strasbourg, France, and
of data products from the Two Micron All Sky Survey, which is a joint
project of the University of Massachusetts and the Infrared Processing
and Analysis Center/California Institute of Technology, funded by the
NASA and the NSF.  We thank T. Willmitch for measuring some of the early KPNO spectra.
The work of FCF and MW has been supported in part by NASA 
grant NCC5-511 and NSF grant HRD-9706268. 
PHASES is funded in part by the California Institute of Technology Astronomy Department, 
and by the National Aeronautics and Space Administration under grant 
NNG05GJ58G issued through the Terrestrial Planet Finder Foundation Science Program. 
This work was supported in part by the National Science Foundation
 through grants AST-0300096, AST-0507590 and AST-005366. MWM is grateful for the 
 support of a Townes fellowship. MK is supported by NASA through grant NNG04GM62G and
the Polish Ministry of Education and Science through 
grant 1P03D 021 29.

\end{document}